\documentclass[%
reprint,
superscriptaddress,
amsmath,amssymb,amsthm,
aps,
soul
]{revtex4-2}

\usepackage{hyperref}
\hypersetup{
  colorlinks   = true, 
  urlcolor     = blue, 
  linkcolor    = blue, 
  citecolor   = blue 
}
\usepackage{algorithm2e}
\SetKwComment{Comment}{/* }{ */}
\RestyleAlgo{ruled}

\usepackage{graphicx}
\usepackage{dcolumn}
\usepackage{bm,soul}
\usepackage{xcolor}
\usepackage{braket}

\usepackage{color}

\begin{document}

\preprint{APS/123-QED}

\title{Parallel Quantum Local Search via Evolutionary Mechanism}

\author{Chen-Yu Liu}
\email{d10245003@g.ntu.edu.tw}
\affiliation{Graduate Institute of Applied Physics, National Taiwan University, Taipei, Taiwan}

\author{Kuan-Cheng Chen}
\email{kuan-cheng.chen17@imperial.ac.uk}
\affiliation{Department of Materials, Imperial College London, London, UK}
\affiliation{Centre for Quantum Engineering, Science and Technology (QuEST), Imperial College London, London, UK}
             
\begin{abstract}
We propose an innovative Parallel Quantum Local Search (PQLS) methodology that leverages the capabilities of small-scale quantum computers to efficiently address complex combinatorial optimization problems. Traditional Quantum Local Search (QLS) methods face limitations due to the sequential nature of solving sub-problems, which arises from dependencies between their solutions.  Our approach transcends this constraint by simultaneously executing multiple QLS pathways and aggregating their most effective outcomes at certain intervals to establish a ``generation''. Each subsequent generation commences with the optimal solution from its predecessor, thereby significantly accelerating the convergence towards an optimal solution. Our findings demonstrate the profound impact of parallel quantum computing in enhancing the resolution of Ising problems, which are synonymous with combinatorial optimization challenges.

\end{abstract}

\maketitle

\section{Introduction}
Leveraging the intricate principles of quantum mechanics, quantum computing emerges as a groundbreaking paradigm, poised to revolutionize problem-solving capabilities far beyond the reach of traditional computing frameworks. This burgeoning field has birthed advanced optimization methodologies, such as the Variational Quantum Eigensolver (VQE) \cite{vqe1} and quantum annealing (QA)\cite{toising1}. These techniques adeptly transform and tackle a wide array of problems through the lens of Ising Hamiltonians or equivalently, as quadratic unconstrained binary optimization (QUBO) \cite{toising2}. The practical applications of these quantum innovations span an impressive spectrum, from optimizing financial portfolios to obtaining solutions for the venerable traveling salesman problem \cite{portfolio1, portfolio2, portfolio3, portfolio4, qaoa1, supplychain1, pqs1, vqc1, rs1, tsp1}.

The advent of the Noisy Intermediate-Scale Quantum (NISQ) era, characterized by quantum systems with a finite number of qubits, necessitates ingenious approaches to fully harness the computational power at our disposal. In this context, the Quantum Local Search (QLS) \cite{qls1} algorithm emerges as a particularly promising strategy. QLS ingeniously circumvents the qubit limitation by iteratively addressing smaller segments of a larger combinatorial puzzle, thereby reducing the quantum resources required for each computation. This segmented approach to problem-solving is versatile, compatible with both gate-based and annealer-based quantum systems. Furthermore, the QLS framework is inherently flexible, allowing for the exploration of various strategies in selecting and tackling sub-problems \cite{rlqls1}, thereby enhancing the overall efficiency and effectiveness of the solution process.

This work proposes an advanced variant of the QLS methodology, termed Parallel Quantum Local Search (PQLS), which ingeniously integrates parallel computing paradigms with quantum computational techniques. PQLS transcends traditional QLS by initiating multiple, concurrent QLS processes, thereby broadening the exploration of the solution space. This parallel approach consolidates the finest outcomes from each strand into a collective ``generation'', leveraging the strengths of each to expedite the journey toward the optimal solution. This innovative methodology not only showcases a marked improvement in solving Ising model-related challenges—a proxy for a wide range of optimization problems—but also symbolizes a significant stride in the utilization of parallel quantum computing to refine and accelerate computational tasks.

\section{Quantum Local Search}

QLS integrates quantum computing with local search methods for combinatorial optimization, aiming to identify optimal solutions within vast, discrete spaces. Unlike classical algorithms that sequentially probe neighboring solutions, QLS employs quantum computers to simultaneously assess multiple candidates, exploiting quantum parallelism for enhanced efficiency \cite{qls1, qbsolv1}.

Consider the Ising model for an optimization problem with $N$ variables:
\begin{equation}
\label{eq:ising}
H = \sum_{i,j=1}^N J_{ij} \sigma_i^z \sigma_j^z + \sum_{i=1}^N h_i \sigma_i^z,
\end{equation}
where $J_{ij}$ and $h_i$ represent coupling and linear terms, respectively. Starting from an initial state $| \psi \rangle$, QLS selects and solves smaller sub-problems using quantum techniques such as VQE, QAOA, and QA, updating the solution based on quantum outcomes. The process iterates until reaching a satisfactory solution or a stop criterion. However, QLS can solve only one sub-problem at a time through its subsequent and iterative updates of local solutions.

\begin{figure*}
\centering
    \includegraphics[scale=0.26]{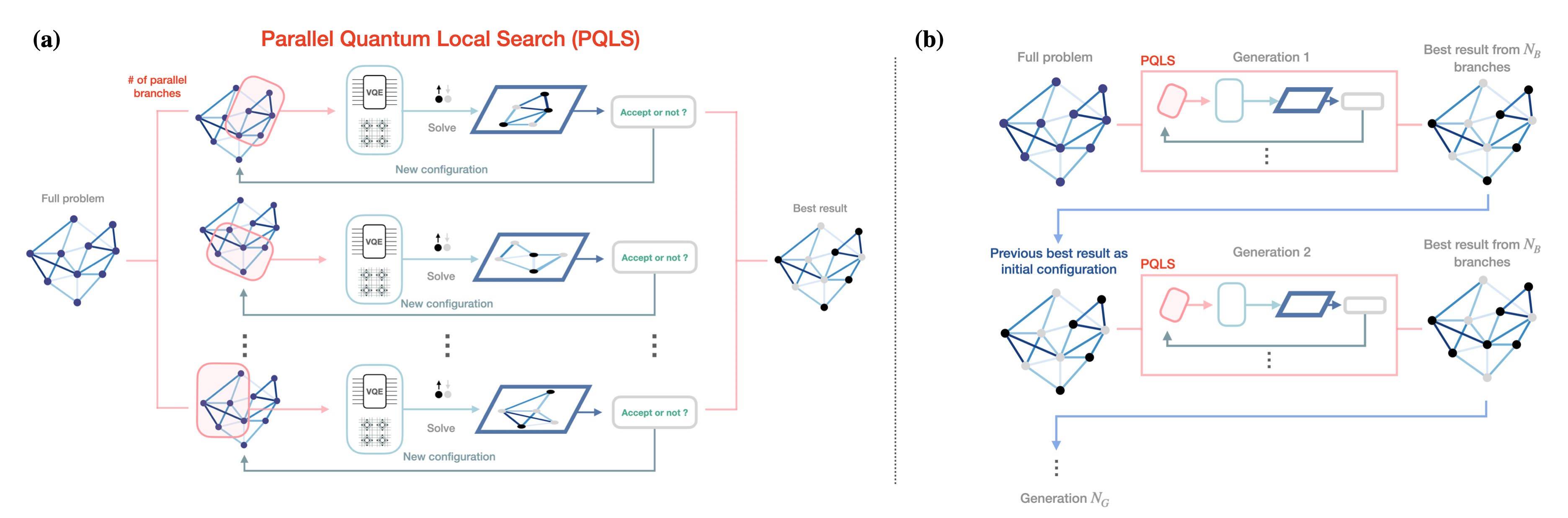}
    \caption{(a) Depicts the foundational structure of PQLS. (b) Expands on the concept by introducing an evolutionary mechanism.
    } 
\label{fig:flow}
\end{figure*}

\section{Parallelization and Evolutionary Mechanism}
We address the limitation of QLS, which can solve only one sub-problem at a time, by proposing a PQLS scheme. As shown in Fig.~\ref{fig:flow}(a), we start with the full Ising problem as defined in Eq.~(\ref{eq:ising}) and create parallel branches of QLS. Each QLS process iterates over a ``branch unit length'' number of iterations. In the end, we obtain the best result among these branches as the output of the PQLS. Currently, the underlying intuition is that this method could indeed yield better results since it expands the search of the solution space in parallel. In fact, this approach can be further enhanced by incorporating an evolutionary mechanism, as illustrated in Fig.~\ref{fig:flow}(b). We design the scheme depicted in Fig.~\ref{fig:flow}(a) to function as a generation cell. This way, the outcome of one generation can serve as the initial solution configuration for the subsequent generation, across a total of $N_G$ generations.

\begin{figure*}
\centering
    \includegraphics[scale=0.27]{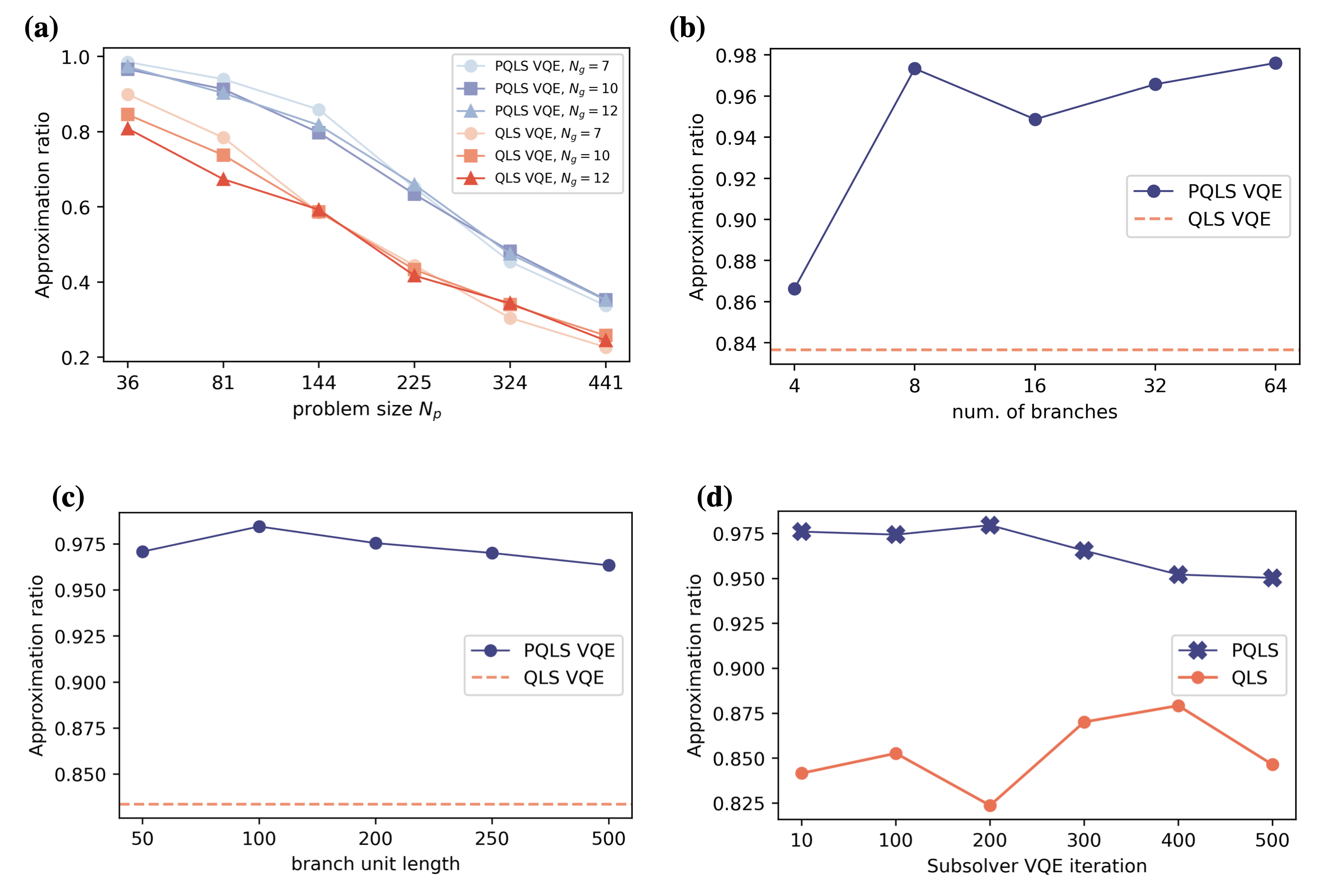}
    \caption{Comprehensive performance analysis of PQLS versus QLS. (a) Comparison across varying problem sizes ($N_p$) and sub-problem sizes ($N_g$) showcases PQLS's superior performance, particularly evident in optimal sub-problem size selections for different $N_p$. (b) Demonstrates the positive impact of increasing the number of branches in PQLS, with a fixed problem size, enhancing solution space exploration. (c) Investigates the influence of branch unit length on PQLS efficiency, identifying an optimal length that maximizes performance, beyond which diminishing returns are observed. (d) Explores VQE iteration variations, highlighting differing optimal iteration counts for QLS and PQLS and underscoring PQLS's enhanced performance due to its parallel processing capabilities.
    } 
\label{fig:result}
\end{figure*}

\section{Result and Discussion}
In Fig.~\ref{fig:result}(a), we present the results of PQLS compared to conventional QLS across various full problem sizes ($N_p$) and sub-problem sizes ($N_g$). The metric ``approximation ratio'' is defined as the ratio between the result obtained and that achieved by the \textsf{Dwave-tabu} solver. For each data point, 5 problems were tested. The branch unit length was set to 100, and the number of branches was 32, with a total of $N_G = 10$ generations. Sub-problems of size $N_g$ were solved using VQE, simulated on IBM quantum platform \cite{ibmquantum}. An interesting trend emerges: for smaller $N_p$ values, the smallest $N_g = 7$ yields the best performance; conversely, for larger $N_p$ values, larger $N_g$ sizes lead to better outcomes. This trend is consistent in both QLS and PQLS; most importantly, PQLS outperforms QLS in every case. 

The results of PQLS with varying numbers of branches are depicted in Fig.~\ref{fig:result}(b). As with the previous analysis, 5 problems were tested for each data point. The total problem size was set as $N_p = 36$, and the sub-problem size was maintained at $N_g = 10$, with a branch unit length of 100 and $N_G = 10$ generations. As expected, an increase in the number of branches leads to better results, since a larger number of branches expands the search space within the solution domain.

Exploring the impact of branch unit length on PQLS performance is crucial. Therefore, in Fig.\ref{fig:result}(c), we adjust the branch unit length from 50 to 500, maintaining settings similar to those in Fig.\ref{fig:result}(b). The number of branches is fixed at 32, and the number of generations is set to $1000/(\text{branch unit length})$. Our findings reveal a noteworthy trend: the performance of PQLS does not increase monotonically with the branch unit length. Instead, there exists an optimal branch unit length that maximizes the algorithm's performance. In the context of our example, this optimal length is identified to be 100. This observation suggests that beyond a certain point, increasing the branch unit length may lead to diminishing returns in solution quality. This could be due to various factors, including the potential for overfitting to specific sub-problems, increased complexity in managing longer branches, or inefficiencies in the use of quantum resources.

The performance of VQE as a subsolver may vary with different iterations. In Fig.~\ref{fig:result}(d), we examine the impact of varying the number of iterations in VQE, which is used to solve sub-problems in both QLS and PQLS. For this analysis, the full problem size is set at $N_p = 36$, the sub-problem size at $N_g = 10$, the number of branches at 64, the branch unit length at 10, and the number of generations at $N_G = 10$. We observe that the optimal number of iterations differs between QLS and PQLS, this variation underscores the different operational efficiencies and problem-solving approaches inherent to the two methodologies. Notably, PQLS demonstrates superior performance across the board, suggesting that its parallel processing capabilities might better exploit the iterative refinement process offered by VQE, leading to more effective solutions. This comparative advantage highlights the importance of iteration tuning in leveraging the full potential of quantum algorithms for complex optimization problems.

\section{Conclusion}
In this work, the proposed PQLS leverages the synergies between parallel computing and quantum mechanics to tackle complex combinatorial optimization problems more efficiently than conventional methods. By executing multiple QLS processes in parallel and optimizing the selection of solutions across generations, PQLS significantly accelerates the convergence toward optimal solutions, as demonstrated through our extensive experimental results.

Our findings highlight the pivotal role of parallelization in quantum computing, especially in the context of the NISQ era, where the limitations of quantum resources necessitate ingenious computational strategies. The comparison between PQLS and conventional QLS across various configurations—such as problem sizes, branch unit lengths, and number of branches—underscores the enhanced efficiency and effectiveness of the parallel approach. Notably, there exists an optimal configuration for these parameters that maximizes performance, indicating the importance of fine-tuning to achieve the best possible outcomes in quantum computational tasks. The involving of an evolutionary mechanism within the PQLS framework opens up new avenues for further research and development. This adaptive strategy, which evolves the solution across generations, promises to enhance the algorithm's capability to navigate the vast solution landscape more effectively, potentially leading to the discovery of superior solutions for even more complex problems. PQLS represents a significant advancement in the field of quantum computing for optimization problems. Its ability to outperform traditional QLS in every tested case not only demonstrates its effectiveness but also serves as a compelling argument for the broader adoption and further exploration of parallel quantum computing strategies in solving real-world challenges. As quantum computing continues to evolve, methodologies like PQLS will play a crucial role in unlocking the full potential of quantum resources, paving the way for groundbreaking solutions across various domains.

\nocite{*}

\bibliography{references}

\end{document}